\documentclass{article}

\usepackage{arxiv}

\usepackage[utf8]{inputenc} 
\usepackage[T1]{fontenc}    
\usepackage[hidelinks]{hyperref}       
\usepackage{url}            
\usepackage{booktabs}       
\usepackage{amsfonts}       
\usepackage{nicefrac}       
\usepackage{microtype}      
\usepackage{amsmath}       
\usepackage{graphicx}
\usepackage[compress]{natbib}
\usepackage{doi}

\title{Can We Replicate Real Human Behaviour Using Artificial Neural Networks?}

\author{ \href{https://orcid.org/0000-0001-9751-0768}{\includegraphics[scale=0.06]{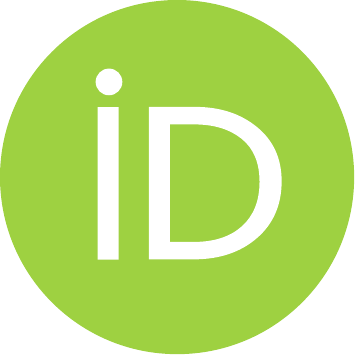}\hspace{1mm}Georg Jäger} \\
	University of Graz\\
	\texttt{georg.jaeger@uni-graz.at} \\
	\And
	\href{https://orcid.org/0000-0002-2541-4467}{\includegraphics[scale=0.06]{orcid.pdf}\hspace{1mm}Daniel Reisinger} \\
	University of Graz\\
	\texttt{daniel.reisinger@uni-graz.at} \\
}

\date{}

\renewcommand{\shorttitle}{Can We Replicate Real Human Behaviour Using Artificial Neural Networks?}

\usepackage{graphicx}
\usepackage{epstopdf}
\usepackage{xcolor}
\bibpunct[, ]{[}{]}{,}{n}{,}{,}
\renewcommand\bibfont{\fontsize{10}{12}\selectfont}
\begin{document}




\maketitle

\begin{abstract}
Agent-based modelling is a powerful tool when simulating human systems, yet when human behaviour cannot be described by simple rules or maximising one's own profit, we quickly reach the limits of this methodology. Machine learning has the potential to bridge this gap by providing a link between what people observe and how they act in order to reach their goal. In this paper we use a framework for agent-based modelling that utilizes human values like fairness, conformity and altruism. Using this framework we simulate a public goods game and compare to experimental results. We can report good agreement between simulation and experiment and furthermore find that the presented framework outperforms strict reinforcement learning. Both the framework and the utility function are generic enough that they can be used for arbitrary systems, which makes this method a promising candidate for a foundation of a universal agent-based model.
\end{abstract}


\section{Introduction}

Agent-based models (ABMs) \cite{abar2017agent,parry2012large,an2012modeling,barbati2012applications} quickly gain importance in various scientific fields \cite{hansen2019agent,lippe2019using,zhang2019empirically,letort2019physiboss,haer2019advancing,geschke2019triple,groff2019state,fagiolo2019validation}. Current crises show us that when predicting the future, human behaviour can be more challenging to model than any natural phenomena, yet it is paramount to the time development of any system involving human beings. Sustainability research is aware of this problem for quite some time now \cite{whitmarsh2012engaging,urry2015climate,beckage2018linking,clayton2015psychological}, but also in the field of virology we found that human behaviour is difficult to predict \cite{chater2020facing,van2020using,betsch2020behavioural} and that agent-based models (ABMs) could help mend this blind spot of many scientific communities \cite{cuevas2020agent,kerr2020covasim,silva2020covid}.

\textcolor{black}{The core idea of ABMs is to shift the focus of a model from equations that try to describe macroscopic properties of a system (e.g. pressure and temperature of a gas) to properties of individual components of the system (e.g. position and velocity of each gas molecule). This way not only information about the macroscopic state of the system can be calculated, but also the states of individual components (i.e. agents) can be observed. The dynamic of the system is then mainly driven by interactions of agents with other agents or their environment. An detailed introduction to the concept of agent-based modelling can, for example, be found in \cite{macal2016everything}.}

There are two factors that limit the wider use of ABMs: First of all, they are difficult to implement, since they usually rely on a set of rules that govern human behaviour, which are difficult to find \cite{klabunde2016decision}. Secondly, and this relates more to using ABMs in scientific communities that are not experienced with this methodology, there is a large gap between the expectation and the capabilities of traditional ABMs: Often the expectation is that ABMs are inherently able to predict human decision making. In reality, describing this process is the most difficult part of designing an ABM \cite{an2020meeting}.

With the advance of machine learning and artificial intelligence \cite{dhall2020machine,sen2020supervised} a solution to this predicament presents itself: We can use an artificial neural network to handle the decision-making of the agents. That way, we do not have to find complicated rules for agent behaviour, rather a utility function that agents want to optimise, using the actions available to them. This combination of methods was used successfully for various different systems \cite{samadi2020decentralized,bone2010simulation,jalalimanesh2017simulation,jang2018agent,hassanpour2021hierarchical,maeda2020deep,collins2014applying,sert2020segregation}. A generic framework that makes use of this synergy was developed in \cite{jager2019replacing} and expanded with an iterative learning approach in \cite{jaeger21}.
\textcolor{black}{Those two studies had the goal of showing that an ABM created from a generic framework using machine learning can lead to the same results as a model that was created manually, thereby shifting the problem of defining rules for agent behaviour to the much simpler problem of finding a utility function. In \cite{jager2019replacing} this was done for the Schelling Segregation model \cite{schelling1971dynamic}, while \cite{jaeger21} investigated the Sugarscape model \cite{epstein1996growing}, a system that is much more difficult to handle for machine learning, since the important states of the system cannot be reached by random decisions of the agents. However, this problem was solved by using an iterative learning approach.}
Furthermore, it was discovered that the framework also has some unique advantages that mainly relate to the fact that a trained neural network has more similarities to a human brain than hard-wired rules. Specifically, the ability to give different agents different experiences proves extremely helpful in modelling human beings \cite{jager2019replacing, jaeger21}.

Previous studies tried to replicate existing models, like the Schelling Segregation model \cite{schelling1971dynamic} or the Sugarscape model \cite{epstein1996growing}. In this work we will try to go one step beyond this and replicate actual human behaviour, extracted from a real-life experiment. The challenge here is, of course, that every human has different experiences and goals and that they do not always act in an optimal way. Rather, they act based on their experience and on certain values they hold, like altruism \cite{fehr2003nature,altru}, fairness \cite{McAuliffe2017} or conformity \cite{confo,rome}. However, exactly for this reason, machine learning is a promising candidate to tackle this challenge.

 The paper is organised as follows: Section \ref{sec:framework} gives a short introduction into the used modelling framework. In Section \ref{sec_util} we construct a generic utility function, so that agent behaviour is more realistic than a simple homo economicus approach. We then use the framework and the utility function to model a public goods game in Section \ref{sec:pgg}. Results are compared to experimental findings and discussed in Section~\ref{sec:results}. Section~\ref{sec:conc} concludes by recapping the most important findings of this study and putting them into context.

\section{Methods}

\subsection{The framework}
\label{sec:framework}

The basic ABM-framework used in this study is detailed in \cite{jaeger21}, and the explanations given here will closely follow this description. The main goal of the framework is to provide a generic and universal tool for developing ABMs.
Like in any other ABM, we have to define what agents can sense, and what actions are available to them. However, instead of defining all the rules that govern the agents' behaviour, we simply have to define one utility function that the agents want to optimize. For most systems, this is a significantly simpler problem. Think of a game of chess: Finding explicit rules on how agents should act given the position of each chess piece is nearly impossible. However, we know that they want to win the game, that they can observe all pieces on the board and which moves they could perform. The heavy lifting of finding the link between observation and optimal action to reach the goal is then performed by a machine learning algorithm. This is closely related to reinforcement learning, yet there are some relevant differences \cite{jaeger21}. In detail, the framework has four distinct phases:

\begin{enumerate}
\item Initialisation
\item Experience
 \item Training 
 \item Application\\
\end{enumerate}

In the initialisation phase we have to define the agents, what they can sense, what they can do and what goal they want to reach, in the form of a utility function.

In the experience phase agents make random decisions and save their experiences. One experience consists of what they sensed, what action they took and what result the evaluation of their utility function yielded. Note, that the agents do not have direct access to their utility function, so initially they are unaware of any possible connections between their utility and the information they can observe. \textcolor{black}{In that sense, the utility function is used as a black box: The agents know the result of the utility function, but do not know how it relates to the variables they can observe. And more importantly: The agents cannot observe all the variables that are needed to calculate the utility function. Only through the use of the artificial neural network they can try to estimate the value of the utility function based on what they observe. Note, that different agents can therefore learn different behaviour, based on what they experienced.}

In the training phase the data gathered in the previous phase are used to train the neural network. \textcolor{black}{Each agent has their own set of experiences and their own neural network.}  In order to keep the framework versatile, we use a hidden layer approach \cite{hansen1990neural,krizhevsky2012imagenet} utilizing cross validation \cite{krogh1995neural}. The artificial neural network is implemented using scikit-learn \cite{pedregosa2011scikit}. \textcolor{black}{We use a Multi-layer Perceptron regressor \cite{murtagh1991multilayer} with a hidden layer consisting of 100 neurons and an adaptive learning rate.} Its goal is to solve the regression problem of predicting the utility of every possible action in the current situation. For a large class of systems it was shown that an iterative approach is necessary for successful training \cite{jaeger21}, in order to avoid the sampling bias that is caused by taking random actions. In such a case, agents go back to the experience phase and make their decisions not randomly, but using their trained neural network.

In the application phase, the agents (now trained and making 'smart' decisions) can be used for modelling purposes and observed as in any conventional ABM.

\subsection{Replicating Human Behaviour}

The overall goal of this study is to bridge the gap between expectation and capability of ABMs by providing a universal framework for ABMs that can replicate human behaviour. For this, we will need to find a generic utility function, that can be used for many different systems. There are certain requirements for that function:

\begin{itemize}
    \item it should only include observables that are present in most systems
    \item it should include properties that can be different for each individual
    \item it should be able to describe human behaviour beyond homo economicus \cite{GINTIS2000311,thaler2000} by including values such as fairness and altruism
\end{itemize}

The first requirement is the strictest, since it means that we can only use variables, that should be present in any ABM. Since ABMs are extremely divers this list is short: All agents have some actions available to them and a goal that they want to reach. Everything else is highly dependent on the system itself. Since also the actions are system-dependent, we can only use the fact that the agents have some goal for constructing a utility function. For most systems, the goal, score or reward of agents is formulated as a payoff, similar to the payoff used in Game Theory. This payoff can now be used as a basis for our utility function. Note the difference: the payoff describes the score or reward of a single agent and maximising this payoff leads to behaviour of a homo economicus. In contrast, the utility function should go beyond that. When thinking about relevant observables related to the payoff, an agent could use their own payoff, the payoff of others, both relative or absolute. This leads to the observables presented in Table \ref{tab1}.

\label{sec_util}

\begin{table}[h]
\center
\caption{Possible observables that can be used in a generic utility function}
\bigskip

\begin{tabular}{|c|c|c|}
\hline
 & \textbf{ absolute} & \textbf{relative} \\ 
 \hline
\textbf{self}   & my payoff   & my payoff / payoff of others \\ 
\hline
\textbf{others} & payoff of others & uniformity of payoffs
\\\hline
\end{tabular}

\label{tab1}
\end{table}

Interestingly, each of these observables can be linked to a personal value that governs human behaviour, as shown in Table \ref{tab2}.

\begin{table}[h]
\caption{Personal values that relate to the relevance of the observables from Table \ref{tab1}}
\bigskip
\center
\begin{tabular}{|c|c|c|}
\hline
 & \textbf{ absolute} & \textbf{relative} \\ 
 \hline
\textbf{self}   & self interest   & conformity \\ 
\hline
\textbf{others} & altruism & fairness
\\\hline
\end{tabular}

\label{tab2}
\end{table}

This will be the basis of constructing the utility function. We define self interest ($si$), altruism ($al$), conformity ($co$) and fairness ($fa$) as values between 0 and 1, with the following meaning:
\textcolor{black}{
\begin{itemize}
    \item $si$ is the payoff the agent wants to receive, given in relation to the maximally possible payoff
    \item $al$ is the payoff the agent wants others to receive, given in relation to the maximally possible payoff
    \item $co$ is the proportion (see Eq.(\ref{dif})) between own payoff and average payoff of others, the agent wants
    \item $fa$ is a weighting factor, that governs how the agent rates the fairness of all payoffs in relation to its own payoff, for states in which all other conditions are met 
\end{itemize}
Additionally we will use $p_s$, the payoff of the agent in relation to the maximally possible payoff and $p_o$, the average payoff of the other agents in relation to the maximally possible payoff.\\
Mathematically, the utility function consists of three main cost terms, i.e. conditions that need to be met, and an additional reward term that helps in comparing two states in which the cost terms yield the same result.
The first term is the self interest cost $C_{si}$. It is zero if the agent gets at least the payoff it wants to receive, otherwise there is a cost linearly connected to the gap between received payoff of the agent $p_s$ and required payoff $si$:
\begin{equation}
C_{si}=max(0,si-p_s)    
\end{equation}
The next cost term is the altruism cost $C_{al}$. It is constructed in a similar way as the self interest cost, but related to the average payoff of other agents $p_o$ and the agents altruism $al$:
\begin{equation}
C_{al}=max(0,al-p_o)
\end{equation}
The final cost term is the conformity cost $C_{co}$. It is zero, if the payoff of the agent $p_s$ is close enough to the average payoff of other agents $p_o$. The allowed gap is given by an agent's conformity $co$. If the gap is too big, i.e. the agents payoff is either too low or too high, the resulting cost is linearly dependent on the size of the gap. For the mathematical definition of this cost term, it makes sense to define a proportion term $prop$ as
\begin{equation}
prop = \frac{min(p_s,p_o)}{max(p_s,p_o)} \label{dif}
\end{equation}
which yields $1$, if $p_s=p_o$ and a non-negative term $<1$ otherwise. With these definitions, the conformity cost $C_{co}$ can be expressed as 
\begin{equation}
C_{co} = max(0,co-prop) \label{dif2}
\end{equation}
Besides these three cost terms, which will decrease the utility, we also need additional terms that increase the utility, if the agent has a high payoff or if the distribution of the payoff is fair, i.e. has a lower Gini coefficient $gc$ \cite{gini,MILANOVIC199745}. The weighting between these somewhat contradictory conditions is given in the agent's fairness $fa$. Thus, we can define a reward function $rew$ as
\begin{equation}
rew = fa \ (1-gc) +  (1-fa) \  p_s
\end{equation}
That means that agents with a fairness of 0 only gain utility from their own payoff, while agents with fairness 1 only gain utility for a low Gini coefficient.\\
When combining the cost terms with the reward term it is important to find a weighting that ensures that the cost terms are the dominant contributions. Since no term can be larger than 1 by definition, this can easily be ensured by multiplying a factor $\lambda$ to the cost terms, similar to a Lagrangian multiplier. Here a factor of 10 was chosen, meaning that even the maximal reward $rew = 1$ can only compensate for a single cost term of $0.1$. Thus, the complete utility function reads:
\begin{equation}
    U=-\lambda \ \left(C_{si} + C_{al} + C_{co}\right) + rew \label{eq1}\\
\end{equation} 
with
\begin{align*}   
    C_{si}&=max(0,si-p_s) \\
    C_{al}&=max(0,al-p_o)\\
    C_{co}&=max(0,co - prop)\\
    rew&= fa \ (1-gc) +  (1-fa) \  p_s
\end{align*}
}

We have now constructed a generic utility function that satisfies the conditions given above and that enables us to try to replicate human behaviour in an ABM using the presented framework.

\subsection{Testing the Framework using a Public Goods Game}
\label{sec:pgg}
In order to test the capabilities of the framework, we will take a look at a system, in which human behaviour is quite counter-intuitive and obviously not only governed by maximising one's own payoff, namely, the public goods game \cite{Axelrod1390}. There are various versions of the public goods game \cite{McGinty2013,fehr99,Andreoni03}, yet they all rely on the same principles: Each individual of a group is given a certain endowment in the form of monetary units. Each individual can then decide for themselves how much (if any) of this endowment they want to contribute to the public good and which part they want to keep for themselves. All contributions in the public good get multiplied by an enhancement factor and then distributed evenly among all individuals, independent of their own contribution. This setup leads to a social dilemma: The maximal public good can be reached when everybody invests everything, yet the maximal personal payoff is reached when a person invests nothing, independent of the contributions of others. In game theory the action of not investing anything is called a dominant strategy, since it yields the best result in all circumstances. Thus, game theory predicts that each person should choose this action.

However, human decision making is not that simple. Many other factors influence one's decision like fairness \cite{fehr2000fairness,bolton2008self} or altruism \cite{Dietz1654,rand14,FEHR2004784}. When observing real people in a setting similar to a public goods game one often finds evidence of cooperation. One of the most prominent examples is a study performed by Fischbacher et al. \cite{fischbacher2001people} in which an experiment is performed to test how people behave in a simple public goods game. The experimental setup was sophisticated so that it was possible to extract what investment people would choose given the contribution of others, without having to play the public goods game over several rounds in order to avoid learning effects or effects of reciprocation.

Fischbacher et al. found, that even though all participants understood the rules of the game, there were three different strategies. The simplest strategy, termed free-riding, is to never contribute anything to the public good. Furthermore, they found behaviour called conditional cooperation, i.e. investing an amount similar to the other group members, but slightly biased towards the own payoff. The third strategy was called hump-shaped, since people following this strategy behave similar to conditional cooperators for overall low contribution, but they start to invest less and less if the contribution of other is higher. This gives us an ideal opportunity to see if we can replicate those strategies using the presented framework.

We will start by defining what actions are available to our agents. We will use the same values as Fischbacher et al., so the agents can contribute anything from 0 to 20 monetary units. Also what they can observe is clear: In the experiment they were informed about what the other group members invested on average, so this is what we will use as sensory input in the model. The payoff $p_s$ is then simply given as
 \bigskip
 
\begin{equation}
 p_s = 20 - inv_s + f \ \frac{\sum_i inv_i}  { N } ,    
 \bigskip
\end{equation}

with the own investment $inv_s$, the sum of all investments $\sum(inv_i)$, the enhancement factor $f = 1.4$ and the number of group members $N$.
\textcolor{black}{Note, that $p_s$ can also be calculated from the average investment of others $inv_{avg}$, which is the sum over all investments minus the investment of the agent itself divided by $N-1$:  
\bigskip
\begin{equation}
 p_s = 20 - inv_s + f \ \frac{\left(inv_s + (N-1) \ inv_{avg}\right)}{N}  
  \bigskip
\end{equation}}
As utility function, we will use Equation (\ref{eq1}), as derived in the previous section.

According to the steps of the framework we will now train the agents using random decisions. An iterative form is possible, but not required, since all the possible states of the system can, in principle, be reached using random decisions.

In the application phase we now try to reproduce the experiment performed by Fischbacher et al. \cite{fischbacher2001people}. In the model, this can be done in a straightforward way, since we can test how agents would behave, given a hypothetical investment of the other agents, without using this data for training, therefore circumventing complications like reciprocity.

Naturally, the results of this experiment will be dependent on the agent's values (self interest, conformity, altruism and fairness) that we choose for the agents. Therefore, we will first ignore conformity and fairness and focus on the interplay between self interest and altruism, scanning this parameter space qualitatively. After that we will try to find reasonable assumptions for those values and see if we can replicate the results obtained by Fischbacher et al.

\textcolor{black}{In order to use the given utility function for modelling purposes, the utilisation of a neural network is absolutely necessary. Since the agents do not have access to all the data required for evaluating the utility function, conventional optimisation techniques fail. The Gini coefficient, needed for evaluating the reward term of the utility function, is an example for such missing information. Since only the average investment of the other agents is known, there is no way of obtaining information about the distribution of investments and therefore the distribution of payoffs. Furthermore, many terms of the utility function depend on the behaviour of the other agents rather than the agent's own behaviour. This makes the optimisation problem exceptionally complex: The agent can only choose its own action and the influence of this action also depends on the personal values of other agents and what they have learned. However, using a neural network offers the possibility to grasp such complex correlations and makes the attempt of modelling this system feasible.}

\section{Results}
\label{sec:results}

\begin{figure}[h] 
\includegraphics[width=\textwidth]{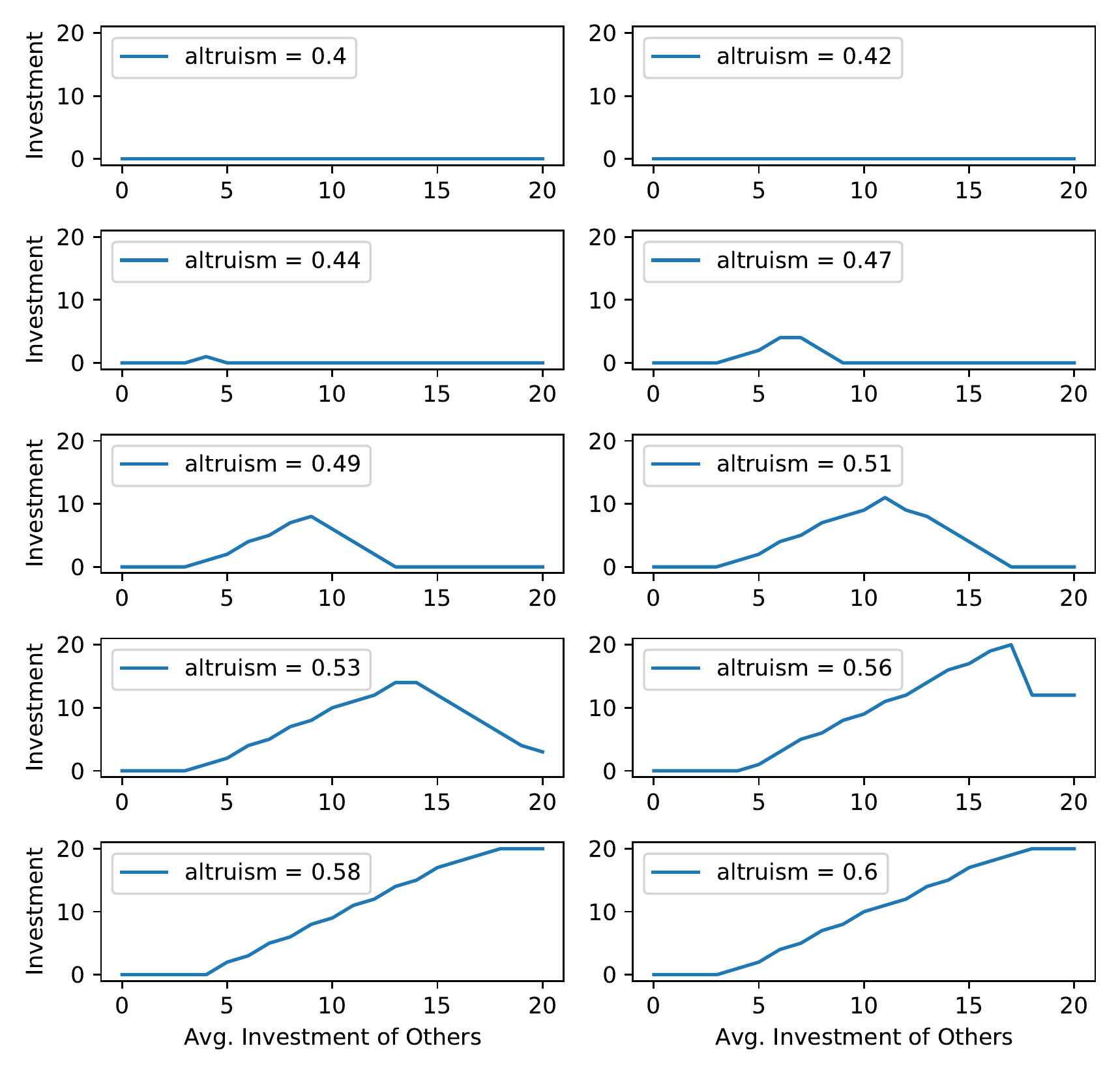}
\caption{Agent behaviour for different values of altruism. The y-axis shows the contribution of the agent, the x-axis shows the average contribution of all other agents.}
\label{fig_res1}
\end{figure}

To see what qualitative behaviour is possible within the framework we will perform a parameter sweep for the value altruism. We keep self interest constant at $0.5$ and vary altruism between $0.4$ and $0.6$. For each value, we will train an agent and see how it reacts, given the contribution of others. Results of this simulation are presented in Figure~\ref{fig_res1}\textcolor{black}{, showing the average investments of 10 agents with identical personal values.}

We find that all three major strategies observed by Fischbacher et al. are present: For low altruism we see free-riding, for medium altruism we see a hump and for high altruism we observe conditional cooperation.

Now that we know that the utility function we found is in principle capable of producing all observed behaviour we can try to find reasonable assumptions for the three strategies and see how well they match the experimental observations. The easiest strategy to emulate is free-riding. \textcolor{black}{Since the goal of free-riding is to optimise one's own payoff, we can simply choose $si = 1$ and keep the other values at 0, meaning that the only relevant goal of the agent is to obtain the maximally obtainable payoff.} For the hump-shaped behaviour we can use a mix of altruism and self interest. Setting both of these variables to $0.5$ means that the agent wants 50\% of the maximal profit both for itself as well as for other agents. If these conditions are met, solutions with higher payoff for the agent itself are preferred, since $fa = 0$. For modelling conditional cooperation we can use not only altruism, but also fairness or conformity. Here we will use $co = 0.8$,
\textcolor{black}{meaning that agents are satisfied as long as their payoff is not more than 20\% lower than average and the average payoff is not more than 20\% lower than their own payoff (see Eq. (\ref{dif2})).} 

\begin{figure}[h] 
\center
\includegraphics[width=0.5\textwidth]{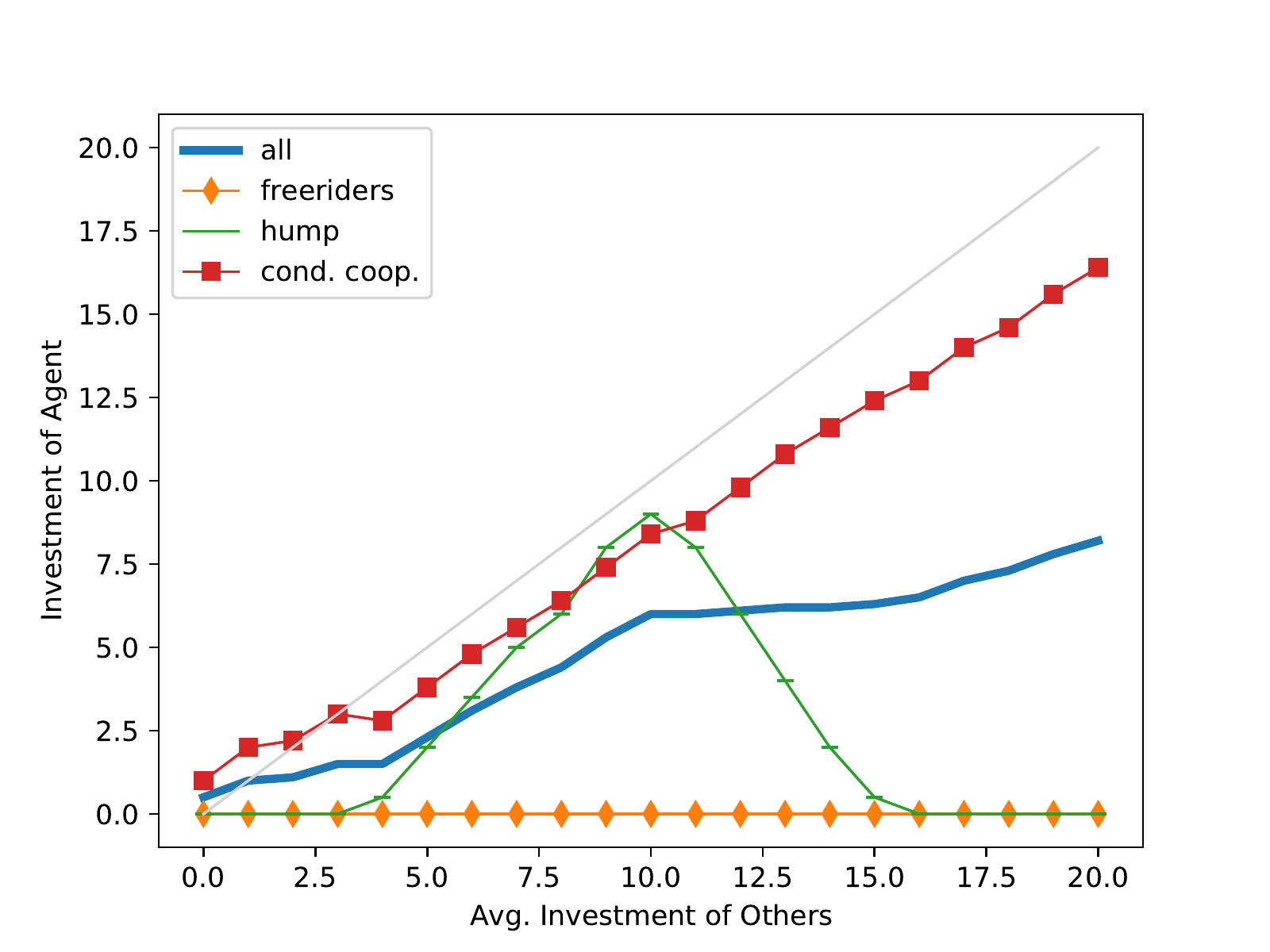}\includegraphics[width=0.5\textwidth]{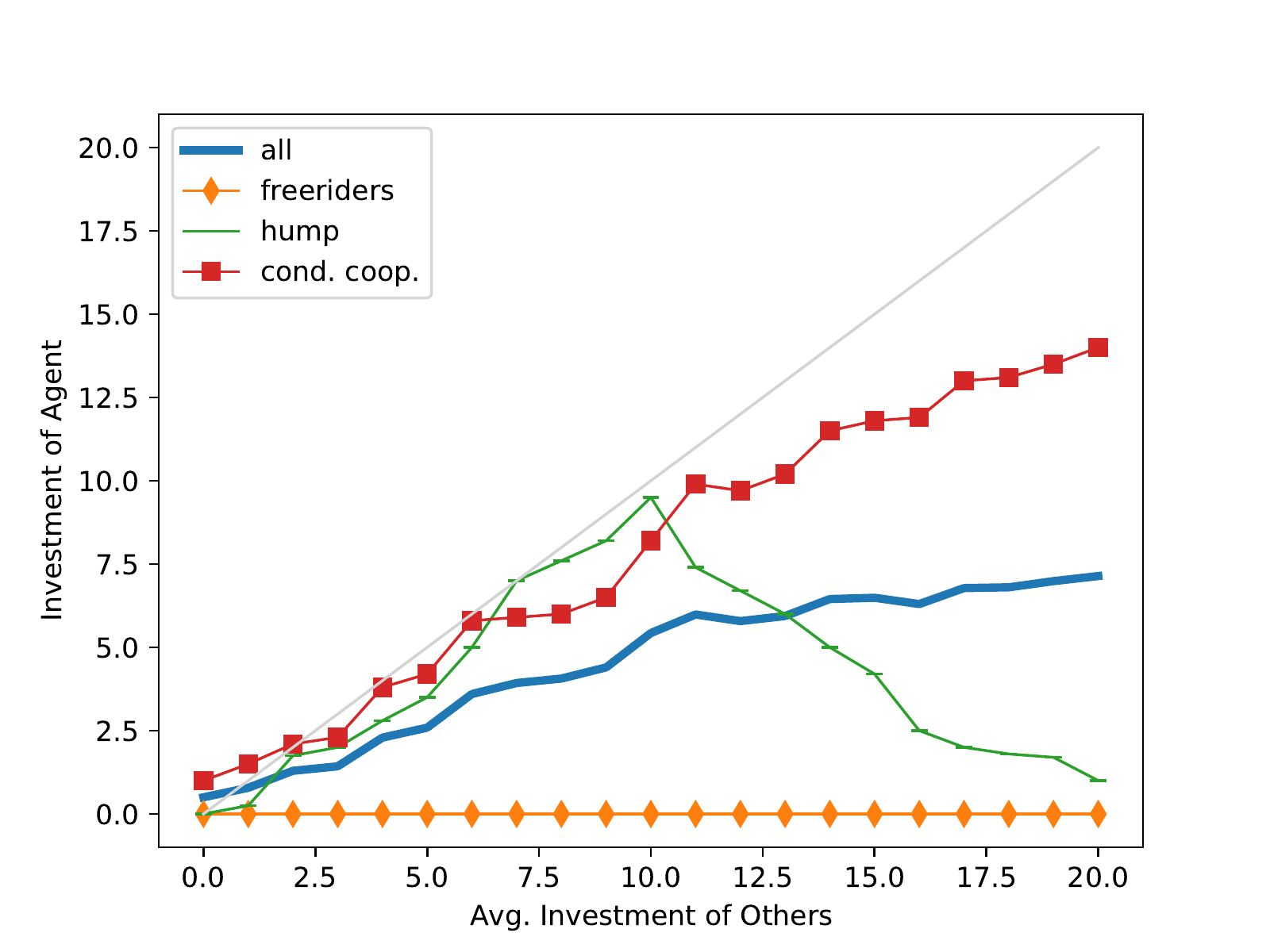}
\caption{Agent behaviour in the simulation (left) and in the experiment (right)}
\label{fig_res2}
\end{figure}

Using these values we try to replicate the experiment performed by Fischbacher~et~al. We train agents using the values determined above and compare their investments with the experimental results. \textcolor{black}{For this, we model an agent population of the same size and strategy composition as the original experiment and attribute them personal values as explained above. After a training phase, in which they play the public goods game, the agents then face the same question as in the experiment: What would they invest, given an average investment of $x$? While in the real experiment this process was rather intricate to ensure that people respond honestly and uninfluenced, in the model it is much simpler: We can observe directly how an agent behaves, given the average investment of others and compare this behaviour to the behaviour of the participants of the original experiment.} This comparison can be seen in Figure~\ref{fig_res2}. The left side shows the simulation results, while the right side shows the experimental results from \cite{fischbacher2001people}.

We find that all three strategies are replicated at least qualitatively. Note, that the numbers chosen for the personal values were not determined by any form of optimisation, but rather by reasonable assumptions. That means, that it would be possible to bring the simulation even closer to the experiment by fine-tuning all parameters. However, this was not the goal of this study. We wanted to show, that, even without fine-tuning, we can reproduce human behaviour patterns based on simple assumptions. 

Beyond the qualitative similarity between experiment and simulation, we can also see some quantitative matches: The hump-shaped strategy has a maximal investment of 10 units for an average investment of 10 units for both simulation and experiment. Also the maximal investment of conditional cooperators and agents overall are a close match.

\begin{figure}[h] 
\center
\includegraphics[width=0.5\textwidth]{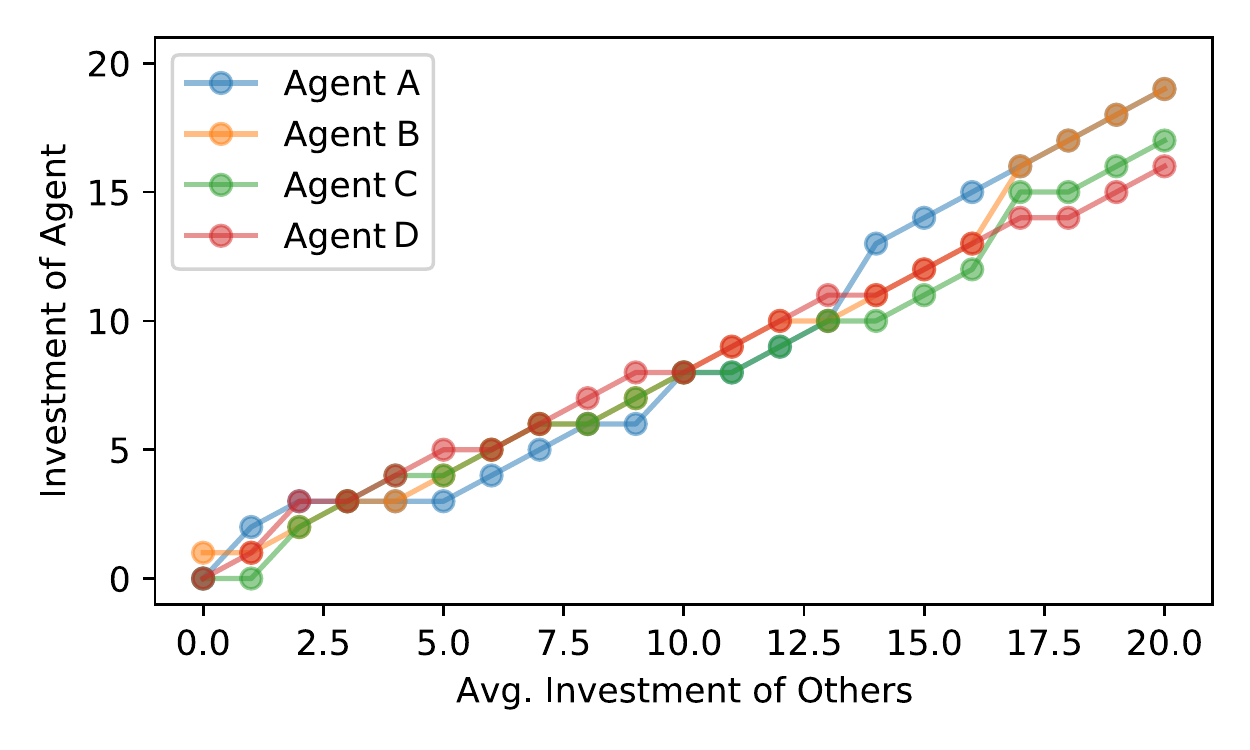}\includegraphics[width=0.5\textwidth]{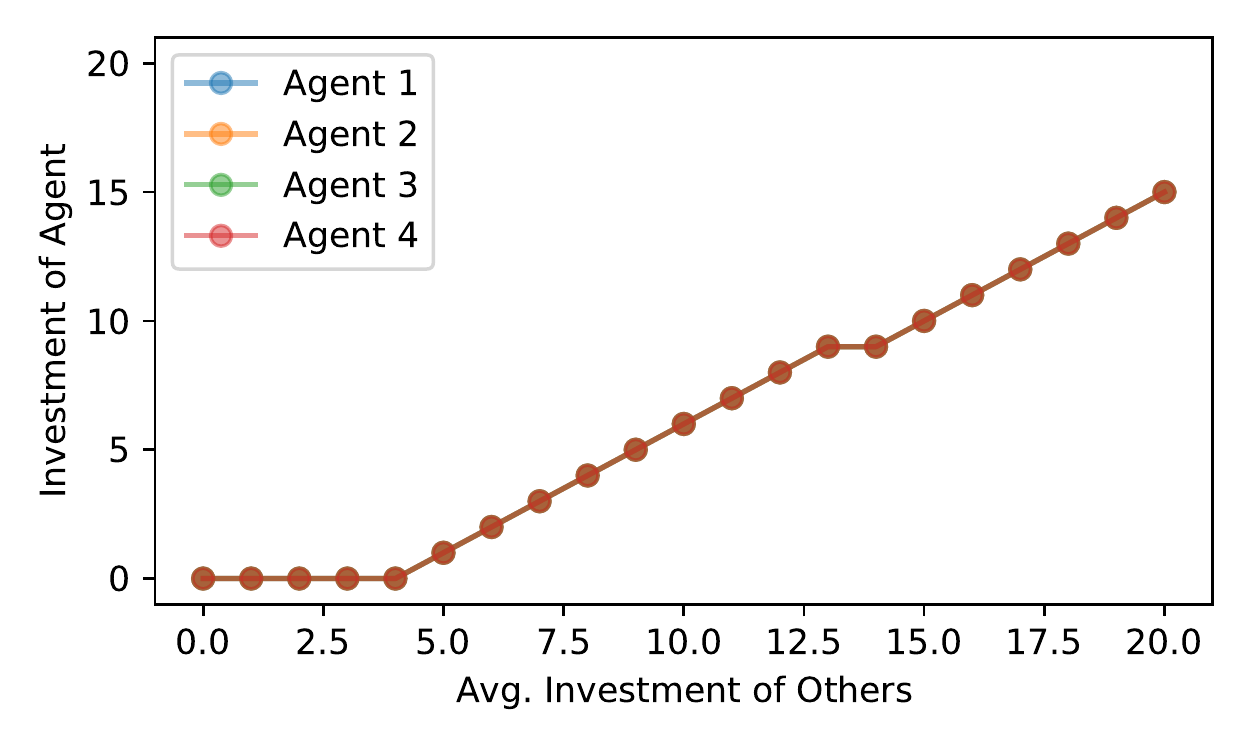}
\caption{The difference between the presented framework (left) and strict reinforcement learning (right)}
\label{fig_res3}
\end{figure}

Note, that this result is different to what would be gained when using strict reinforcement learning. The goal of reinforcement learning is to get an optimal result, i.e. to give the agents all the possible information as well as the function they need to optimise and then train until they make perfect predictions. In contrast, here we only provide the information humans would have access to and do not provide the utility function itself to the agents. Training is performed until a certain prediction accuracy is reached, but this threshold is chosen below 100\%, in order to do justice to the fact that people have no accurate, quantitative understanding of their own utility function. \textcolor{black}{Here, this threshold is chosen at 99.0\%, but for 95\% - 99.9\% the obtained results look qualitatively similar. Note, that using the prediction accuracy of the neural network is only one possibility of modelling the inaccuracy of human predictions. While other techniques, like introducing noise to the input or output of the neural network may give more control, the utilisation of the prediction accuracy has a significant advantage: Any kind of (white) noise averages out over the course of training and will be ignored or averaged by the agents, having little influence on their behaviour outside the training phase. Only by using the prediction accuracy to stop the training at a given time we can make lasting changes to the trained network and therefore the behaviour of the agents. The inclusion of this human inaccuracy drastically improves the realism of the framework. This can be seen in Figure \ref{fig_res3}, which compares the results of the framework with the results of traditional reinforcement learning.} 

Here we see the behaviour of four agents with the same personal values ($co = 0.8$, i.e. conditional cooperation). The left side depicts their decisions when this ABM framework is used, the right side is the result of reinforcement learning. 
We see the two important differences: Using reinforcement learning leads to the same behaviour for all agents, since they can exactly calculate the effect of each of their actions, independent of the experiences they gathered. Secondly, we notice some form of risk aversion: In the left panel, agents choose an investment of which they are sure will satisfy their need for conformity, while the agents in the right panel choose the absolute minimal investment that meets this requirement.

\section{Conclusion}
\label{sec:conc}
We used a framework for agent-based modelling based on machine learning to model a public goods game. Contrary to the traditional approach, we did not use the payoff of each agent as the sole determinant of an agents goal, but rather a universal utility function that includes the agents own payoff and the payoff of others to calculate a reward or score for this agent, taking into account its personal values. In order to make this function generic, we considered the relevance of the own payoff and the payoff of others both in a relative and in an absolute way, leading to the four values self interest, conformity, altruism, and fairness, defined in Section \ref{sec_util}.

We then compared our simulation results with the experimental results of Fischbacher et al. \cite{fischbacher2001people} and found good agreement. This means that the presented framework and utility function is a good candidate for modelling various systems from game theory and beyond. We also found that the framework outperforms strict reinforcement learning in terms of realistic behaviour of the agents (see Figure \ref{fig_res3}). Even when we use exactly the same utility function, reinforcement learning leads to rigid behaviour, while the framework can produce agents with the same values that still act differently because they gathered different experiences. This leads to a much more realistic description of human behaviour.

Regarding applicability, the framework can be used for most multi-agent systems. However, if the rules of behaviour are quite clear and simple to formulate there is no real benefit of using the framework, as the traditional ABM approach leads to the same behaviour. However, for systems in which\\

\begin{itemize}
    \item the decision process is complex,
    \item parts of the relevant information is shrouded from the agents, or
    \item agents have different values and goals,
\end{itemize}

 using the framework offers a significant advantage. Still, for systems in which the goal of the agents is completely unknown the framework cannot be used in a straightforward way. Nevertheless the framework could be used to test different ideas for goals, values or utility functions to see which of those produce the observed behaviour, similar to using traditional ABMs to revers engineer rules for behaviour \cite{gilbert2000build,van2012agent}.

 An interesting expansion for the framework would be including network effects of social networks. In the current implementation, every agent can observe any other agent (as was the case in the experiment to which we compared our results). However, in a larger system people can only observe someone who is in some way connected to them. Using the link strength as weighting factor in the utility function would even enable us to give higher importance to the well-being of close relatives than to casual acquaintances. The possibility of negative link strength could even describe some kind of rivalry, which would further improve the realism produced by the framework. While such an expansion would add nothing substantial to this study, it would be an interesting way of further expanding the framework's scope.

One of the main goals of this framework was to keep it generic enough to work for nearly all systems, that can be modelled using an ABM. Both the framework itself as well as the utility function presented here satisfy these demands. However, it is still not clear if this formulation of the utility function is the optimal choice for all systems. We showed that it succeeds in replicating realistic human behaviour in the public goods game, yet more testing is necessary, before the framework can be released as open source code. Reaching that goal, however, would be a significant step for agent-based modelling, since it would give the scientific community a flexible tool for creating models that can realistically depict human behaviour, even in research communities that are not not yet familiar with the method of agent-based modelling.




\bibliographystyle{unsrt}
\bibliography{nnabm}

\end{document}